\documentclass[11pt,a4paper]{article}
\usepackage{amsfonts}
\usepackage{amsmath}
\usepackage{amsthm}
\usepackage{amssymb}
\usepackage{amscd}

\def\be{\begin{equation}}
\def\ee{\end{equation}}

\def\bea{\begin{eqnarray}}
\def\eea{\end{eqnarray}}

\def\tth{\tilde{\theta}}
\def\tph{\tilde{\phi}}
\def\tps{\tilde{\psi}}

\begin{document}

\begin{flushright}
DAMTP-2001-111
\end{flushright}

\vspace{2cm}
\begin{center}
{\LARGE {\bf Axisymmetric non-abelian BPS monopoles from $G_2$ metrics}}
\vspace{1cm}

Sean A. Hartnoll \\
\vspace{0.3cm}

{\it DAMTP, Centre for Mathematical Sciences, Cambridge University,\\
Wilberforce Road, Cambridge CB3 0WA, UK.}
\vspace{0.5cm}

\noindent e-mail: S.A.Hartnoll@damtp.cam.ac.uk
\end{center}
\vspace{1cm}

\begin{abstract}
Exact $SU(2)\times U(1)$ self-gravitating BPS global monopoles in four dimensions are constructed by
dimensional reduction of eight dimensional metrics with $G_2$ holonomy asymptotic to cones
over $S^3\times S^3$. The solutions carry two topological charges in an interesting way.
They are generically axially but not spherically symmetric. This last
fact is related to the isometries and asymptotic topology of the $G_2$ metrics.
It is further shown that some $G_2$ metrics known numerically
reduce to supersymmetric cosmic strings.
\end{abstract}

\pagebreak
\pagenumbering{arabic}

\section{Introduction}

Self-gravitating non-abelian solitons (\cite{vg} for a review) are natural generalisations of flat space
solitons. From the original Bartnik-McKinnon solutions \cite{bm} onwards, interesting
behaviour has been found in static Einstein-Yang-Mills systems.
For example, non-abelian black holes can violate many no-hair and uniqueness theorems
\cite{vg}. However, not many exact self-gravitating non-abelian static solutions are known. One such solution
\cite{cv1, cv2}, of $\mathcal{N}=4$ gauged supergravity, has been used recently to
construct a supergravity dual to large $N$, $\mathcal{N}=1$ super Yang-Mills \cite{mn}.

Recent work on manifolds with $G_2$ holonomy \cite{bs,gpp,cglp1,cglp3,bggg,cglp4,cglp5,cglp6,ab} provides an easy way to obtain
supersymmetric solitons by dimensional reduction. Manifolds of $G_2$ holomony are seven dimensional
manifolds that admit a parallel spinor. They 
are therefore important in supersymmetric compactifications of eleven dimensional supergravity or M-theory.
Here we will trivially extend the Riemannian $G_2$ manifolds to eight dimensions by adding a time direction and then
dimensionally reduce to obtain supersymmetric monopole solutions of a four dimensional theory.

Four families of noncompact $G_2$ manifolds asymptotic to cones over $S^3\times S^3$ are known, denoted
$\mathbb{B}_7, \mathbb{C}_7, \widetilde{\mathbb{C}}_7, \mathbb{D}_7$ in \cite{cglp6}. The metrics
all have an isometry group containing $SU(2)\times SU(2) \times U(1)$.
We will dimensionally reduce on $SU(2)\times U(1)$ contained in the
$SU(2)\times SU(2)$. This will result in static four dimensional manifolds with an $SU(2)/U(1) = S^2$
factor and $U(1)$ isometry,
corresponding to axially symmetric monopole or cosmic string solutions. There will also
be $SU(2)\times U(1)$ gauge fields and scalars. The Bogomol'nyi equations have effectively already
been solved in constructing the $G_2$ metric and we will see that the four dimensional solutions
are automatically BPS. This method allows
the construction of exact supersymmetric solutions that would be hard to guess directly in four dimensions.
The correspondence between special holonomy metrics in higher dimensions and BPS solutions
has been used recently in the other direction in \cite{e} in which $SU(2)$ instantons on
$S^4$ were shown to give give rise to Spin(7) metrics on the chiral spinor bundle of $S^4$.

The dimensionally reduced Lagrangian density in four dimensions will be \cite{ss}
\bea\label{eq:lag}
\mathcal{L} & = & - R - \frac{1}{4} e^{\phi} F^{a \mu \nu} F^b_{\mu \nu} \Phi_{a b} + \frac{1}{4} g^{\mu \nu}
\mathcal{D}_{\mu} \Phi_{a b} \mathcal{D}_{\nu} \Phi^{a b} - \frac{1}{2} g^{\mu \nu} \partial_{\mu} \phi
\partial_{\nu} \phi \nonumber \\
& & - \frac{1}{4} e^{-\phi} \left[-4 \Phi^{i i} + \epsilon_{ijk}\epsilon_{lmn} \Phi_{il} \Phi^{jm}\Phi^{kn}\right] ,
\eea
where we have written $e^{2\phi} \equiv \det \Phi$ to emphasise the dilatonic behaviour of the
determinant. The Higgs fields $\Phi_{a b}$ are in the second symmetric power of the adjoint
representation of the gauge group $SU(2)\times U(1)$, transforming as (\ref{eq:symm}) below. This is
somewhat uncommon in the context of monopoles but arises naturally in dimensional
reduction \cite{ss}. The Lagrangian has standard
Eintein-Yang-Mills-Higgs(-dilaton) terms with a potential that is unbounded below.
Here and throughout, indices $a,b,...$ run from $1$ to $4$ and indices $i,j...$ run from $1$ to $3$.
The metric convention is $(-,+,+,+)$.

The terms in the Lagrangian (\ref{eq:lag}) come from dimensional reduction of the Einstein-Hilbert action
in eight dimensions. The full supersymmetric theory will have fermionic fields and also more bosonic
fields coming from an eight dimensional supergravity (e.g. \cite{sase}). These other fields are set to zero in
the solutions discussed here.

Section 2 reviews the relevant results on $G_2$ metrics. Section 3 is the dimensional reduction from eight to
four dimensions. Section 4 discusses the metric, gauge fields and scalars. It is seen that the solutions
correspond to global monopoles as they are asymptotically conical.
They are generically not spherically symmetric.
This result is discussed in the context of previous work on non-spherically symmetric black holes \cite{rw1, rw3, rw2, hkk}
and is related to recent work on the asymptotic topology of $G_2$ metrics \cite{cglp3, cglp4, ab}.
It is shown that although defining magnetic charges by integration of field strengths
over spheres at infinity is problematic, nontrivial topological charges may be associated with the scalar fields.
Section 5 considers some aspects of four dimensional solutions corresponding
to more general $G_2$ metrics than considered in the previous
sections. Solutions corresponding to cosmic strings are found. Section 6 is the conclusion.

\section{$G_2$ metrics asymptotic to cones over $S^3 \times S^3$}

Three families of complete nonsingular seven dimensional
metrics with $G_2$ holonomy are known, based on generalisations of \cite{bs, gpp}.
They are asymptotic to cones over $\mathbb{CP}^3$, $SU(3)/T^2$ and $S^3 \times S^3$, and hence
noncompact. These topological spaces belong to a restricted set of possibilities for
cohomogeneity one $G_2$ metrics \cite{cs}. The last of these cases has
isometries appropriate for dimensional reduction to a self-gravitating non-abelian monopole in four dimensions.

A general ansatz for the $G_2$ metric with nine radial functions was introduced in \cite{cglp1}.
\be \label{eq:9fun}
ds_7^2 = dr^2 + a_i(r)^2 (\tilde{\sigma_i} + g_i(r) \sigma_i)^2 + b_i(r)^2 \sigma_i^2 ,
\end{equation}
where $\sigma_i$ and $\tilde{\sigma_i}$ are the left invariant one forms on each copy of $SU(2) = S^3$. That is,
\bea\label{eq:1forms}
\sigma_1 = \cos\psi d\theta + \sin\psi \sin\theta d\phi  & \tilde{\sigma_1} = & \cos\tps d\tth + \sin\tps \sin\tth d\tph, \nonumber\\
\sigma_2 = - \sin\psi d\theta + \cos\psi \sin\theta d\phi & \tilde{\sigma_2} = & - \sin\tps d\tth + \cos\tps \sin\tth d\tph, \nonumber\\
\sigma_3 = d\psi + \cos\theta d\phi & \tilde{\sigma_3} = & d\tps + \cos\tth d\tph.
\eea
The ranges for the coordinates are $0 \leq \theta,\tth \leq \pi$, $0 \leq \phi,\tph < 2\pi$ and
$0 \leq \psi,\tps \leq 4\pi$.
The metric (\ref{eq:9fun}) generically has isometry group $\widetilde{SU(2)} \times SU(2)$ corresponding
to left multiplication of $SU(2)$ on each of the $S^3$s. 
The condition for $G_2$ holonomy becomes a set of nine first order equations for $a_i, b_i, g_i$.
The general solution to these coupled nonlinear equations is not known. This
is one reason to consider consistent truncations of the nine function ansatz by setting various of the
radial functions to be equal. Another reason is that
an extra $U(1)$ isometry is needed to give the M-theory solution a Kaluza-Klein interpretation in terms of type
IIA string theory.

Various truncated solutions involving six radial functions, providing the desired $U(1)$ isometry,
have been studied and a unified description has recently been given \cite{cglp5,cglp6,ab}. Four families of
solutions are known numerically, denoted $\mathbb{B}_7, \mathbb{C}_7, \widetilde{\mathbb{C}}_7, \mathbb{D}_7$.
A generic member of any family is aymptotically locally conical (ALC) as opposed to asymptotically
conical (AC), meaning that there is an $S^1$ that stabilises as $r\to\infty$. This will be the orbit of the
$U(1)$ isometry. All four families of metrics have a limiting case in
which they become AC. The families have different behaviours at the origin.
Thus $\mathbb{B}_7$ and $\mathbb{D}_7$ have an $S^3$ bolt whilst
$\mathbb{C}_7$ and $\widetilde{\mathbb{C}}_7$ have a $T^{1,1} = (S^3 \times S^3) / S^1$ bolt. A bolt
is a subspace that remains of finite size in a degenerate orbit at the origin,
the principal orbits here are $S^3\times S^3$.
For the cases with an $S^3$ bolt, the AC limit
corresponds to the original $G_2$ metric of \cite{bs,gpp}
and for this metric the $U(1)$ isometry is enhanced
to a third $SU(2)$, as will be discussed below.
From an M-theory perspective, the most interesting
result is a unified treatment of the deformed - corresponding to $\mathbb{B}_7$ - and resolved -
corresponding to $\mathbb{D}_7$ - conifolds in type IIA string theory \cite{cglp5, cglp6, ab}.

We will concentrate on the few cases in which a closed form solution
is known. Other cases will be considered in \S 5.
This begins with an ansatz for the metric with six radial functions studied in \cite{cglp1, bggg},
\be \label{eq:6fun}
ds_7^2 = dr^2 + a_i(r)^2 (\tilde{\sigma_i} - \sigma_i)^2 + b_i(r)^2 (\tilde{\sigma_i} + \sigma_i)^2 .
\end{equation}
In \cite{cglp3} it was shown that for the case of a collapsing $S^3$ at the origin, i.e. the $\mathbb{B}_7$
family in the notation of the previous paragraph, the only regular solutions of the six-function equations are also
solutions of a reduced set of four-function equations, with the metric ansatz now written as
\bea\label{eq:4fun}
ds_7^2 = dr^2/c(r)^2 & + & a(r)^2 \left[ (\tilde{\sigma_1} - \sigma_1)^2 + (\tilde{\sigma_2} - \sigma_2)^2 \right]
+ b(r)^2 \left[(\tilde{\sigma_1} + \sigma_1)^2 + (\tilde{\sigma_2} + \sigma_2)^2 \right] \nonumber\\
& + & c(r)^2 (\tilde{\sigma_3} + \sigma_3)^2 + d(r)^2 (\tilde{\sigma_3} - \sigma_3)^2. 
\eea
This metric has an $\widetilde{SU(2)} \times SU(2) \times U(1) \times \mathbb{Z}_2$ symmetry,
corresponding to the symmetries of an M-theory lift of $N$ D6-branes wrapping the $S^3$ of the
deformed conifold geometry.

An exact solution of the four-function equations was found to be \cite{bggg}
\bea\label{eq:gener}
a(r) = \frac{\sqrt{(r-r_0)(r+3r_0)}}{\sqrt{8r_0}}, & c(r) & =  \frac{\sqrt{2r_0(r^2-9r_0^2)}}{\sqrt{3(r^2-r_0^2)}}, \nonumber\\
b(r) = \frac{\sqrt{(r+r_0)(r-3r_0)}}{\sqrt{8r_0}}, & d(r) & = \frac{r}{\sqrt{6r_0}}, 
\eea
where $r_0$ is a scale parameter, present for any Ricci-flat metric.
The range of the radial coordinate here is $3r_0 \leq r < \infty$, so it will generally be more
convenient to work with the shifted coordinate $\bar{r} = r - 3r_0$. The corresponding metric is ALC
because $c(r)$ remains finite at infinity.
It was shown \cite{bggg,cglp3} that this solution extends to a two parameter family of solutions, $\mathbb{B}_7$,
that are not known explicitly. A limiting case of this family is the AC solution which is known exactly\footnote{The
AC metric is more commonly expressed in terms of the radial variable $\rho=\sqrt{r}$.}
\bea\label{eq:special}
a(r) & =  d(r) & =  \sqrt{r}, \nonumber\\
b(r) & =  c(r) & =  \sqrt{\frac{r}{3}} \sqrt{1-\left(\frac{r_0}{r}\right)^{3/2}}, 
\eea
where again $r_0$ is a scale parameter and $r_0 \leq r < \infty$, so it will be convenient to use the
shifted coordinate $\bar{r} = r - r_0$. The singular conifold is obtained in the limit $r_0 \to 0$.
The AC metric has an enhanced isometry group
$SU(2)^3 \times \mathbb{Z}_2$. We will see below that the fact that the third term in the isometry group for
the generic ALC metric is $U(1)$ and not $SU(2)$ will mean that the global monopole in four dimensions
is axially but not spherically symmetric. Finally, note that in the AC case, the $\mathbb{Z}_2$ symmetry
can be interpreted as a spontaneously broken ``triality'' symmetry that is important in the physics of M-theory
on the $G_2$ manifold \cite{aw}.

\section{From the $G_2$ metric to the global monopole}

First, trivially extend the $G_2$ metric of (\ref{eq:4fun}) to an eight dimensional Lorentzian manifold
\be\label{eq:8dim}
ds^2_{1,7} = - dt^2 + ds^2_7
\end{equation}
This will be a solution to the eight dimensional Einstein vacuum equations. We want to dimensionally
reduce on an internal group manifold $G = SU(2) \times U(1)$ to get non-abelian gauge fields for $G$,
scalars transforming in an adjoint representation of $G$, and a four dimensional Lorentzian metric.
Part of the isometry group of the metric (\ref{eq:4fun}) is $\widetilde{SU(2)} \times SU(2)$, by action of $SU(2)$
on the two sets of left invariant one forms (\ref{eq:1forms}). We will take the $\widetilde{SU(2)}$ to be part
of the internal group manifold. Then observe \cite{cgllp}
that there are three commuting $U(1)$ Killing vectors of the metric (\ref{eq:4fun}), which
can be taken to be $\frac{\partial}{\partial \phi}$, $\frac{\partial}{\partial \tph}$
and $\frac{\partial}{\partial \psi}+\frac{\partial}{\partial \tps}$. The
last of these cannot be simultaneously reduced with $\widetilde{SU(2)}$ whilst the first two are related by the
$\mathbb{Z}_2$ symmetry of the metric. So we further reduce on the $U(1)$ generated by $\frac{\partial}{\partial \phi}$.
This leaves a remaining $SU(2)/U(1)=S^2$, so we expect there to be an $S^2$ factor in the reduced metric, at
least topologically. Further, we make the indentification $\psi \sim \psi + 2\pi$, to half the range of $\psi$. This is a
symmetry of the metric (\ref{eq:4fun}) without fixed points - the bolt $S^3$ is of finite size at the origin -
on the manifold and so will not introduce orbifold singularities.

The eight dimensional metric had an
unbroken supersymmetry, due to the special holonomy, and the reduction process commutes
with the supersymmetry transformations. We can see this as follows. The isometry group
acts on the parallel spinor to give another parallel spinor. By considering the
supersymmetry transformations, we see that the transformed parallel spinor will
be the same as the original parallel spinor if and only if the isometry transformation
commutes with supersymmetry. But it is a fact that $G_2$ metrics admit only one
parallel spinor \cite{jo}. Therefore supersymmetry commutes with the isometry group and
therefore the dimensional reduction does not ruin supersymmetry. In other words
the lower dimensional solution will have unbroken
supersymmetries, in fact one supercharge, and will be a BPS soliton. It should thus be stable.
The Bogomol'nyi equations for gravitating systems are the first order consistency equations for the
existence of parallel spinors on the background \cite{vg}. To explicitly find the corresponding
generalised BPS bound it is likely that the arguments of \cite{gkltt,gt} will need to be modified
because the spacetime in the present case is not asymptoticaly flat.

To dimensionally reduce, we want to rewrite the metric (\ref{eq:8dim}) in the usual form for
Scherk-Schwarz reductions \cite{ss,dnp}
\be\label{eq:red}
ds^2_{1,7} = (\det\Phi)^{-1/2} ds^2_{1,3} + \Phi_{a b} [ \lambda^a + A^a ]
[ \lambda^b + A^b ] ,
\end{equation}
where the left invariant forms for the internal $SU(2)\times U(1)$ are $\lambda^i = \tilde{\sigma_i}$ and $\lambda^4 = d\phi$.
Note that $\Phi,A,ds^2_{1,3}$ are functions of $(t,r,\theta,\psi)$ whilst dependence on the internal coordinates $(\phi,\tth,\tph,\tps)$
is restricted to the $\lambda^a$.

By comparing (\ref{eq:red}) with (\ref{eq:4fun}) one can read off the scalar fields
{\footnotesize
\be\label{eq:phi}
\Phi_{a b} = \left(\begin{array}{cccc}
a^2+b^2 & 0 & 0 & (b^2-a^2) \sin\psi \sin\theta \\ 
0 & a^2+b^2 & 0 & (b^2-a^2) \cos\psi \sin\theta \\ 
0 & 0 & c^2 + d^2 & (c^2-d^2) \cos\theta \\ 
(b^2-a^2)\sin\psi \sin\theta & (b^2-a^2)\cos\psi \sin\theta & (c^2-d^2) \cos\theta & (a^2+b^2) \sin^2\theta + (c^2+d^2) \cos^2\theta\\ 
\end{array}\right)_{a b} ,
\end{equation}
}
with determinant
\be\label{eq:det}
\det\Phi = 4 (a^2+b^2)^2 c^2 d^2 \left[ (\alpha(\bar{r})-1) \sin^2\theta + 1\right] ,
\end{equation}
and the gauge fields are
\bea\label{eq:gauge}
A^1 & = & - \frac{4 (b^2-a^2) (a^2+b^2) c^2d^2}{\det\Phi} \left[\sin\psi \sin\theta\cos\theta d\psi -\left[\left(\alpha(\bar{r})-1\right)\sin^2\theta + 1\right] \cos\psi d\theta\right],\nonumber\\
A^2 & = & - \frac{4 (b^2-a^2) (a^2+b^2) c^2d^2}{\det\Phi} \left[\cos\psi \sin\theta\cos\theta d\psi
+\left[\left(\alpha(\bar{r})-1\right)\sin^2\theta + 1\right] \sin\psi d\theta\right] ,\nonumber\\
A^3 & = & \frac{4 (a^2+b^2) (c^2-d^2) a^2b^2}{\det\Phi} \sin^2\theta d\psi ,\nonumber\\
A^4 & = & \frac{4 (a^2+b^2)^2 c^2d^2}{\det\Phi}\cos\theta d\psi ,
\eea
where
\be
\alpha(\bar{r}) = \frac{a^2b^2(c^2+d^2)}{c^2d^2(a^2+b^2)} .
\end{equation}
And the four dimensional metric - note that in (\ref{eq:red}) we have already rescaled to be in the Einstein frame - is
\be\label{eq:met4}
ds^2_{1,3} = \det\Phi(\bar{r},\theta)^{1/2} \left( - dt^2 + \frac{d\bar{r}^2}{c^2} + R^2(\bar{r}) d\Omega^2_{\bar{r}} \right),
\end{equation}
where
\be
R^2(\bar{r}) = \frac{4a^2b^2}{(a^2+b^2)} ,
\end{equation}
and $d\Omega^2_{\bar{r}}$ is a metric on the sphere at radius $\bar{r}$,
\be\label{eq:sph}
d\Omega^2_{\bar{r}} = d\theta^2 + \frac{\sin^2\theta}{(\alpha(\bar{r})-1)\sin^2\theta+1} d\psi^2 .
\end{equation}

It is already clear that the special AC metric (\ref{eq:special}) will result in a simpler solution
because in this case $\alpha(\bar{r}) = 1$ and much of the angular
dependence will vanish. For this family of solutions, $\alpha(\bar{r})= 1$ is the
condition for spherical symmetry.

\section{Discussion of metric, scalars and gauge fields}

\subsection{Special case: the metric}

Consider first the AC case (\ref{eq:special}) in which $a(\bar{r})=d(\bar{r})$ and $b(\bar{r})=c(\bar{r})$.
The metric has the expected $S^2$ factor, with coordinates $0 \leq \theta \leq \pi$ and $0 \leq \psi < 2\pi$.
Because $\alpha(\bar{r})=1$ in this special case, the metric on the sphere is the usual round metric.
Furthermore, $\det\Phi = 4(a^2+b^2)^2 a^2b^2$, with no $(\theta,\psi)$ dependence, and therefore
the metric is spherically symmetric.

As $\bar{r} \to 0$, the metric is, up to an overall constant and with $\rho = 4 \bar{r}^{3/4}/3$,
\be\label{eq:speci}
ds^2_{1,3} \sim - \frac{1}{2}\left(\frac{3\rho}{4}\right)^{2/3} dt^2 + d\rho^2 + \frac{9}{16}\rho^2 (d\theta^2 + \sin^2\theta d\psi^2) .
\end{equation}
There is a conical singularity at the origin, hidden as a point of infinite redshift. An infinite redshift at the
origin is seen in other contexts such as \cite{mtw} pg. 683, which is the metric describing a self-similar
star cluster. Interestingly, the star cluster metric is asymptotically conical, which also turns out
to be the case here because as $\bar{r} \to \infty$, the metric is, letting $\rho=2\bar{r}^{3/2}/3$ and up to
an overall constant,
\be \label{eq:metlim}
ds^2_{1,3} \sim - \frac{1}{3}\left(\frac{3 \rho}{2} \right)^{4/3} dt^2
+ d\rho^2 + \frac{3}{4} \rho^2 (d\theta^2 + \sin^2\theta d\psi^2) .
\end{equation}
This metric is asymptotically conical (in the three dimensional sense now) with a deficit solid angle.
Asymptotically conical
metrics are characteristic of {\bf global monopoles} \cite{bv}. In the
manifestly asymptotically conical coordinate system one must have $T_{0 0}
\sim \frac{1}{\rho^2}$, where $\rho$ is the radial coordinate in (\ref{eq:metlim}). Thus the monopole has an
infinite positive energy.

The singular conifold in seven dimensions is obtained as the limit $r_0 \to 0$. Near the origin in this limit, the metric in
four dimensions will just be (\ref{eq:metlim}). So the four dimensional metric will also have a conical singularity
at the origin in this limit.

\subsection{Generic case: the metric}

Consider now the generic ALC case, concentrating on the closed form four-function solution (\ref{eq:gener}).
Again, we have the expected $S^2$ factor.
However, the metric on the sphere is not the round metric. Thus the solution is not spherically symmetric. We have a well defined metric on $S^2$, because as $\theta \to 0$ then $d\Omega^2_{\bar{r}}$ goes as $(1+\mathcal{O}(\theta^2)) d\theta^2 + (\theta^2 + \mathcal{O}(\theta^3))d\psi^2$ for all $\bar{r}$. In (\ref{eq:met4}) note that $\alpha(\bar{r}) \to 1$
as $\bar{r} \to 0$ and $\alpha(\bar{r}) \to \infty$ as $\bar{r} \to \infty$, this implies that the $S^2$ is round near the origin and becomes increasingly stretched and cylindrical in the $z$ direction as we move out radially.

As $\bar{r} \to 0$, the metric takes the following form up to a constant with $\rho = 4\bar{r}^{3/4}/3$
\be
ds^2 \sim - \frac{1}{2}\left(\frac{3\rho}{4}\right)^{2/3} dt^2 + d\rho^2 + \frac{9}{16}\rho^2 (d\theta^2 + \sin^2\theta d\psi^2) .
\end{equation}
The metric is the same as the special case (\ref{eq:speci}), a fact we will see directly in \S 5 below.
Again there is a conical singularity and point of infinite redshift at the origin.
As $\bar{r} \to \infty$, the metric is up to constant and letting $\rho = \bar{r}^3/3$
\be\label{eq:limbig}
ds^2 \sim \sin\theta \left( - (3\rho)^{4/3} dt^2 + \frac{3}{2 r_0}
\left[ d\rho^2 + \frac{9 \rho^2}{6} d\Omega^2_{\rho} \right] \right).
\end{equation}
This metric is not quite asymptotically conical as $d\Omega^2_{\rho}$ has
a dependence on $\rho$ (\ref{eq:sph}). The $S^2$ is increasingly
cylindrical as we move out. This fact and the overall $\sin\theta$ term
means that there is not a straightforward way of defining the energy of the solution.
It is in the same family, $\mathbb{B}_7$, as the special solution we found before,
and could be called by analogy a global monopole.

The most interesting feature of the metric is the lack of spherical symmetry. As mentioned in \S2 above, this can
be understood from the symmetries of the eight dimensional metric. We reduced on
a $U(1) \subset SU(2)\times U(1)$, where the embedding is entirely into the first term of the direct product.
The isometry group of the reduced metric will be the normaliser of
this $U(1)$ subgroup modulo the $U(1)$ itself. This is because the remaining symmetry must commute
with the symmetry we are reducing, otherwise it would not have a well-defined
action on the reduced manifold. We quotient out the $U(1)$, which trivially normalises
itself, because it has been quotiented out of the metric in the reduction.
It is immediately seen that the reduced normaliser is just the second term of the direct
product, i.e. $U(1)$. This is the axial symmetry. In the special case above we had started with $SU(2)\times SU(2)$
in the original metric, the reduced normaliser was then $SU(2) \simeq SO(3)$ and we obtained spherical symmetry.

Lack of spherical symmetry has been discussed before in the context of hairy black holes with
Yang-Mills-Higgs matter. It was shown in \cite{rw1} that a magnetically charged
Reissner-Nordstr\"om black hole embedded in a theory with additional massive charged vector fields
is unstable under perturbations in these vector fields if the horizon radius is less than the radius
of the magnetic monopole core. Physically this is the fact that production
of charged vector particles is energetically favourable in a sufficiently strong magnetic field
because these particles carry magentic moments that can be aligned to partially shield the magnetic field
\cite{rw3}.

In \cite{rw2}, perturbative static solutions away from the Reissner-Nordstr\"om solution
were found with nonzero massive vector fields. As the Reissner-Nordstr\"om metric is
spherically symmetric with magnetic charge $n$, these vector fields can be expanded in monopole vector harmonics.
The total angular momentum of these harmonics cannot be zero unless $n=1$, due to a
contribution to the angular momentum of magnitude $eg=n$ directed along the line from the monopole
to the charged particle. This implies that if $n>1$ then the solution cannot be spherically symmetric.
Recently \cite{hkk}, exact solutions for these axially symmetric black holes have been found numerically and
it was shown that they should be classically stable. In the present situation, the
massive vector fields are due to spontaneous symmetry breaking as we will see below. We have an exact
analytic solution for an axially symmetric metric with Yang-Mills and Higgs fields.

Another result closely related to the present situation is that spherically symmetric
global monopoles of scalar fields have been shown to be at best marginally stable under
certain axially symmetric perturbations \cite{g,au}.

To summarise, for the family of metrics considered here \footnote{We
will see in \S 5 below that this correspondence only holds for the
metrics with an $S^3$ bolt, $\mathbb{B}_7$ and $\mathbb{D}_7$. The
metrics with a $T^{1,1}$ bolt do not have enhanced isometry in the AC
case.} we have that AC $\leftrightarrow$ spherical symmetry and
ALC $\leftrightarrow$ axial symmetry.
A result on $G_2$ metrics reviewed in \S2 was that such metrics asymptotic to cones over $S^3 \times S^3$
were generically ALC. A result reviewed in the preceeding paragraphs is that spherically
symmetric (global) monopoles with charged massive vector fields may not be stable under certain axial pertubations.
This suggests that one might be able to understand the preference for ALC metrics by lifting
results about instabilities of spherically symmetric monopole configurations to the higher dimensional metrics.

\subsection{Generic case: the scalars and gauge fields}

Does the solution carry topological charges? This will depend firstly upon how much gauge symmetry is broken by the scalar
fields. The scalars are in the second symmetric power of the adjoint representation of the gauge group so they transform as
\be\label{eq:symm}
\delta_{\xi}\Phi_{a b} = f_{a c d} \xi_c \Phi_{d b} + f_{b c d} \xi_c \Phi_{a d} ,
\end{equation}
where $f_{a b c}$ are the structure constants. In our case $G = SU(2) \times U(1)$ so $f_{i j k} = \epsilon_{i j k}$ and all
other components are zero. Symmetries are preserved if $\delta_{\xi}\Phi_{a b} = 0$, which then allows
costruction of a gauge invariant field strength $\xi^a F^a$ \cite{ad}.

The situation is subtle, because although scalar fields transforming in adjoint representations are usually
interpreted as Higgs fields associated with symmetry breaking, here they couple to the gauge fields and
metric in such a way that
$\det\Phi$ behaves rather like a dilaton coupling $e^{2\phi}$. So one needs to disentange `dilaton' and
`Higgs' type behaviour of these fields. The scalar potential, in the Einstein frame Lagrangian \cite{ss}, is
\be
U(\Phi) = \frac{1}{4} \frac{1}{(\det\Phi)^{1/2}} \left[-4 \Phi^{ii} +
\epsilon_{ijk}\epsilon_{lmn} \Phi_{il} \Phi^{jm}\Phi^{kn}\right] ,
\end{equation}
which is unbounded below (the internal space was not Ricci-flat) and without critical points and so
it is at first unclear how to define the
vacua. This is characteristic of dilaton potentials. Correspondingly, the scalar fields (\ref{eq:phi}) diverge at infinity.
However, if we normalise the
scalar fields to make them all finite at infinity by dividing by $\frac{1}{\bar{r}^2}$ then we can consider them to
belong to a vacuum moduli space corresponding to minimising the `Higgs' part of the potential.

Inserting (\ref{eq:phi}) as $\bar{r} \to \infty$ divided by $\frac{1}{\bar{r}^2}$ into (\ref{eq:symm}) one finds that
the $SU(2)$ symmetry is broken at infinity to a $U(1)$, with symmetry generator $\xi = (0,0,\epsilon,0)$. This depends
crucially on the fact that in this normalisation $\Phi_{1 4} \sim \Phi_{2 4} \to 0$ as $r \to \infty$ because
$a^2-b^2 \to \infty$ slower than $a^2, b^2$ or $d^2$.
Further, the remaining $U(1)$, generated by $\xi^{\prime} = (0,0,0,\epsilon)$, is unbroken throughout all of space.
Thus there are potentialy {\it two} magnetic charges.

A naive attempt to define the magnetic charge fails. As $\bar{r} \to \infty$ then $F^4 \sim \frac{1}{\bar{r}^4}$, and
similarly the gauge invariant field strength of the $U(1) \subset
SU(2)$ goes as $F \equiv F^a \xi^a \sim \frac{1}{\bar{r}^4}$.
Therefore we cannot form a magnetic charge by integrating over the sphere at spatial infinity. This is reminiscent of
the situation for gravitational sphalerons \cite{bm, gv} which are unstable. However, the probable stability of the present
solution means that either there is a topological charge with
the usual relationship between the magnetic integral at infinity
and the charge not holding or the situation is similar to the
Chamseddine-Volkov
solution, which doesn't have any Higgs fields and
hence no topological charge \footnote{The Chamseddine-Volkov solution \cite{cv1,cv2} also does not, in fact, have a
gauge invariant magnetic charge defined by an integral at infinity. Its stability will presumably follow along the lines
of \cite{fg}.}. A good indication that the former
possibility is the case here is that the usual asymptotic relationship
relating the gauge and scalar fields for monopoles,
$D_{\mu} \xi^a  = 0$, does not hold for this solution.

Consider first the $U(1)$ not coming from the $SU(2)$. Because the symmetry is unbroken
throughout space, we can consider the field near the origin $\bar{r} \to 0$ and find a potential,
in cartesian coordinates,
\be
A^4 \sim \frac{z}{\sqrt{x^2+y^2+z^2}} \frac{(x dy - y dx)}{x^2+y^2} .
\end{equation}
This is clearly the potential corresponding to a Dirac magentic monopole with charge $m=2$. And this charge is
topological in nature and so is the charge of this $U(1)$.

For the other $U(1)$ we need to look
at the homotopy class of the map from $S^2_{\infty}$ to the moduli space of Higgs vacua which is
$G/H$, where $H$ is the unbroken subgroup of the gauge group $G$. Here this is $SU(2)/U(1) = S^2$.
This can be topologically nontrivial because for $G$ simply connected $\pi_2(G/H) \cong \pi_1(H) \cong \mathbb{Z}$.
We can find the degree of the map $S^2 \to S^2$ in two ways. For the first method, decompose the matrix
of scalar fields into various $SU(2)$ representations
\be\label{eq:decom}
\frac{1}{\bar{r}^2}\Phi = \left(\begin{array}{cc} 
{\bf A} & {\bf v} \\ 
{\bf v}^t & s \\ 
\end{array}\right) ,
\end{equation}
where ${\bf A}$ is 3 by 3 matrix, ${\bf v}$ is a 3 component vector and $s$ is a scalar. The matrix can be visualised
as an ellipsoid, defined by its three eigenvalues and the vector gives an oriented direction. So from (\ref{eq:phi})
\be
{\bf A} = \left(\begin{array}{ccc}
k_1 & 0 & 0  \\ 
0 & k_1 & 0  \\ 
0 & 0 & k_2  
\end{array}\right) ,
\end{equation}
and
\be
{\bf v} = (-\frac{1}{\bar{r}} k_3 \sin\theta \cos\psi, -\frac{1}{\bar{r}} k_3 \sin\theta \sin\psi, -k_2\cos\theta) .
\end{equation}
In these expressions, $k_1,k_2,k_3$ are positive constants. We see that the ellipsoid ${\bf A}$ breaks the $SU(2) \simeq SO(3)$ 
symmetry to rotations about one axis, i.e. $U(1) \cong SO(2)$. Then the vector ${\bf v}$ will completely break the symmetry,
unless it is pointing along the axis of symmetry of the ellipsoid. And this is precisely what happens as $\bar{r} \to \infty$ and the first
two components vanish! This gives a geometric understanding of the partial symmetry breaking at infinity found above. The vacua
are thus defined by an alligned ellipsoid and vector. And hence just by the direction of the vector (the normalisation is
unimportant). Thus the map $S^2 \to S^2$ is
\be\label{eq:wind}
(\theta, \psi) \mapsto \lim_{\bar{r} \to \infty} \frac{1}{\sqrt{\frac{k_3^2 \sin^2\theta}{\bar{r}^2} + k_2^2\cos^2\theta}} (-\frac{1}{\bar{r}} k_3 \sin\theta \cos\psi, -\frac{1}{\bar{r}} k_3 \sin\theta \sin\psi, -k_2\cos\theta)
= (0,0,\pm 1) ,
\end{equation}
with the positive value for $\theta > \frac{\pi}{2}$ and negative for $\theta < \frac{\pi}{2}$. At $\theta = \frac{\pi}{2}$
the vector is zero, so the direction is determined by the ellipsoid, which is ambiguous up to a $\mathbb{Z}_2$ flip in direction.
This map has
{\bf degree $1$}. This is easiest seen by considering $\bar{r}$ as defining a homotopy of maps
with $\bar{r}$ going from $\bar{r}=k_3/k_2$, which is just the identity map from the sphere to the sphere with degree one, to
$\bar{r} \to \infty$ which is the map we are interested in. The degree is homotopy invariant so the map in (\ref{eq:wind})
has degree one.

Alternatively, we can construct the moduli space of vacua explicitly by acting with $G=SU(2)$
on a particular vacuum element that we know from our solution
\be\label{eq:vac}
\Psi_0(0,0) = \left(\begin{array}{cccc}
k_1 & 0 & 0 & 0 \\ 
0 & k_1 & 0 & 0 \\ 
0 & 0 & k_2 & -k_2 \\ 
0 & 0 & -k_2 & k_2 \\ 
\end{array}\right) ,
\end{equation}
where $k_1, k_2$ are fixed positive constants, and without loss of generality
we are considering the point $\theta=0$. Varying $\theta$ must keep us within the space of vacua, after
a suitable normalisation.
Strictly the action is of $SU(2)\times U(1)$ but we have already shown
that the $U(1)$ leaves these fields invariant. The most general nontrivial finite transformation is by a group
element $g = e^{-\alpha T_1} e^{-\beta T_2}$. Where $\{T_a\}$ are the generators of the adjoint representation
of $SU(2)\times U(1)$.
Under this transformation, (\ref{eq:vac}) becomes
{\small
\bea
\Psi_0(\alpha,\beta) =
k_1 \left(\begin{array}{cccc}
\cos^2\beta + \sin^2\alpha\sin^2\beta & \cos\alpha\sin\alpha\sin\beta
& -\cos^2\alpha\cos\beta\sin\beta  & 0 \\ 
\cos\alpha\sin\alpha\sin\beta & \cos^2\alpha &
\cos\alpha\sin\alpha\cos\beta & 0 \\ 
-\cos^2\alpha\cos\beta\sin\beta & \cos\alpha\sin\alpha\cos\beta
& \sin^2\beta + \sin^2\alpha\cos^2\beta & 0 \\ 
0 & 0 & 0 & 0 \\ 
\end{array}\right) \nonumber \\
+
k_2 \left(\begin{array}{cccc}
\cos^2\alpha\sin^2\beta & -\cos\alpha\sin\alpha\sin\beta
& \cos^2\alpha\cos\beta\sin\beta  & - \cos\alpha\sin\beta \\ 
-\cos\alpha\sin\alpha\sin\beta & \sin^2\beta &
-\cos\alpha\sin\alpha\cos\beta & \sin\alpha \\ 
\cos^2\alpha\cos\beta\sin\beta & -\cos\alpha\sin\alpha\cos\beta
& \cos^2\alpha\cos^2\beta & -\cos\alpha\cos\beta\\ 
-\cos\alpha\sin\beta & \sin\alpha & -\cos\alpha\cos\beta & 1 \\ 
\end{array}\right) .
\eea
}
This gives us the full moduli space of vacua, $S^2(\alpha,\beta)$, with $0 \leq \alpha \leq \pi, 0 \leq \beta < 2\pi$.
Now we need to find the degree of the map $S^2_{\infty} \to S^2(\alpha,\beta)$ given by
$\alpha(\theta,\psi),\beta(\theta,\psi)$ such that
\be\label{eq:map}
\lim_{\bar{r} \to \infty} \frac{1}{\bar{r}^2} \Phi(\bar{r},\theta,\psi) = \Psi_0(\alpha(\theta,\psi),\beta(\theta,\psi)) .
\end{equation}
Note that there is in fact no dependence on $\psi$ in the case under consideration. In this equation,
the positive constants $k_i$ will in general be different in $\Phi$ and $\Psi_0$.
The solution to (\ref{eq:map}) is given by
\bea
\alpha & = & 0 \mbox{  if  } \theta < \frac{\pi}{2} , \nonumber \\
\alpha & = & \pi \mbox{  if  } \theta > \frac{\pi}{2} , \nonumber \\
\beta & = & 0 .
\eea
Which is exactly what we found in (\ref{eq:wind}) above.

So the enhanced gauge symmetry at infinity, in which the $SU(2)$ is not completely broken, does in fact
result in a topological charge. This is clearly related to the possibility of defining a gauge invariant magnetic
charge, even though this vanishes.

\subsection{Special case: the scalars and gauge fields}

The special spherically symmetric case similarly has two topological
charges. Furthermore, one can also define corresponding
charges by integrals at spatial infinity.

Consider first the $U(1)$ not in the $SU(2)$. The gauge field is just $A^4 = \cos\theta d\psi$ which is exactly
the potential for a Dirac monopole with magnetic moment $m=2$. Because this is now true as $\bar{r} \to \infty$,
the topological charge arises in the usual fashion and coincides with the integral of $F^4 = dA^4 \sim \frac{1}{\bar{r}^2}$
over the sphere at spatial infinity.

Consider now the $U(1) \subset SU(2)$. With the notation of (\ref{eq:decom}), one has for the scalar fields,
\bea
{\bf A} & = & (a^2+b^2) {\bf I}_3 , \nonumber\\
{\bf v}^t & = & (b^2-a^2) (\sin\psi \sin\theta, \cos\psi \sin\theta,\cos\theta) .
\eea
In the geometrical language of the previous subsection, the ellipsoid is in fact a sphere and therefore
the spontanteous symmetry breaking is due solely to the vector {\bf
v}. The generator of the unbroken $U(1)$ is clearly
$\xi = \epsilon (\sin\psi \sin\theta, \cos\psi \sin\theta,\cos\theta,0)$. Because the ellipsoid does not
contribute to the symmetry beaking, the gauge symmetry is everywhere
broken to $U(1)$. The
corresponding gauge invariant field strength is 
\be
F \equiv \xi^a F^a = \xi^a (d A - A \wedge A)^a = -3 \frac{4r^3-r_0^3}{(4 r^{3/2}-r_0^{3/2})^2} d\theta \wedge \sin\theta d\psi .
\end{equation}
This can be integrated to give a finite magnetic charge
\be
Q_B = \int_{S^2_{\infty}} F = -3\pi .
\end{equation}
The existence of a corresponding topological charge
is seen as the unit winding number of the indentity map $S^2 \to S^2$, exactly as for the usual
Yang-Mills-Higgs monopole.

\section{Monopoles and cosmic strings from other $G_2$ metrics}

So far we have only considered reductions of the $\mathbb{B}_7$ family of $G_2$ manifolds, and only
the two cases which have a known closed form solution for the radial functions. However, Taylor
series expansions about the origin are known for all the families of metrics discussed in \S 2
\cite{cglp3,cglp4,cglp5,cglp6,ab} and these can be used to examine the behaviour of the reduced
four dimensional metrics near the origin. Further, the asymptotic behaviour of the radial functions
can be found from their first order equations and this allows us to study the asymptotics of the
four dimensional metrics.
All of these seven dimensional metrics have a $U(1)$ isometry in the generic
ALC case and therefore will have axial symmetry. If they did not have
the $U(1)$ isometry, the reduced metric would not even be axially
symmetric. We will see now that the cases with an $S^3$ bolt (collapsing $S^3$) are global monopoles and the
cases with a $T^{1,1}$ bolt (collapsing $S^1$) are cosmic
strings. These are all solutions of the same four dimensional theory (\ref{eq:lag}).

\subsection{$\mathbb{B}_7$ and $\mathbb{C}_7$}

The metrics for the $\mathbb{B}_7$ and $\mathbb{C}_7$ families with a $U(1)$ symmetry both have the form
\bea
ds_7^2 & = & dr^2 + a_1(r)^2 \left[ (\tilde{\sigma_1} - \sigma_1)^2 + (\tilde{\sigma_2} - \sigma_2)^2 \right] \nonumber \\
& + & b_1(r)^2 \left[ (\tilde{\sigma_1} + \sigma_1)^2 + (\tilde{\sigma_2} + \sigma_2)^2\right]
+ a_3(r)^2 (\tilde{\sigma_3} - \sigma_3)^2
+ b_3(r)^2 (\tilde{\sigma_3} + \sigma_3)^2 .
\eea
The Taylor expansions of the radial functions about the origin, to the order that we need them, are for $\mathbb{B}_7$ \cite{cglp3}:
\bea
& & a_1(r) \sim a_3(r) \sim 1 + \frac{1}{16} r^2 , \nonumber \\
& & b_1(r) \sim b_3(r) \sim -\frac{1}{4} r ,
\eea
where a trivial scale factor has been fixed to one. We see that an $S^3$ collapses at the origin. For $\mathbb{C}_7$ the
Taylor expansions are \cite{cglp4}:
\bea
a_1(r) \sim 1 - \frac{q}{8} r + \frac{(16-3q^2)r^2}{128} , & & a_3(r) \sim -r , \nonumber \\
b_1(r) \sim 1 + \frac{q}{8} r + \frac{(16-3q^2)r^2}{128} , & & b_3(r) \sim q + \frac{q^3 r^2}{16},
\eea
where a trivial scale factor has again been fixed to one and $q$ is a constant. Regularity
requires $\mid q \mid \leq q_0 = 0.91\cdots$ with $\mid q \mid = q_0$ corresponding to the AC solution.
We see that now an $S^1$ collapses at the origin.

The dimensional reduction is now done as in \S 3. We are interested here in the behaviour of the spatial
metric near the origin, so we will not consider the time component. For the $\mathbb{B}_7$ case one
obtains up to a constant, after putting $\rho = 2 r^{3/2}/3$,
\be\label{eq:b7}
ds^2_3 \sim d\rho^2 + \frac{9 \rho^2}{16} (d\theta^2+\sin^2\theta d \psi^2) .
\end{equation}
This is just the metric at the origin that we found in \S 4. Note that the $r$ here is a different
radial coordinate to that used above. For $\mathbb{C}_7$ one obtains up to a constant
\be\label{eq:c7}
ds^2_3 \sim \sin\theta (dr^2 + 2 d\theta^2 + 4r^2 d\psi^2) .
\end{equation}
The metric has become cylindrical, with axial not spherical symmetry
at the origin. Further, there is a surfeit plane angle
suggesting that the solution corresponds to a {\bf cosmic
string}. This can be understood as the fact that in the $G_2$ metric
only an $S^1$ collapses at the origin.
Note that this solution remains axially not spherically
symmetric for all allowed values of $q$. In particular the AC limit
does not restore spherical symmetry.

The asymptotic behaviour as $r\to\infty$ of the radial functions can be found from the first order equations they
satisfy. From the equations in \cite{cglp3,cglp4} one has for both $\mathbb{B}_7$ and $\mathbb{C}_7$ that
{\it in the ALC case}
\bea
a_1(r)^2 \sim \frac{r^2}{12} , & & a_3(r)^2 \sim \frac{r^2}{9} , \nonumber \\
b_1(r)^2 \sim \frac{r^2}{12} , & & b_3(r)^2 \sim k^2,
\eea
where $k$ is a constant. Doing the dimensional reduction and setting
$\rho = r^3/3$ the metric up to a constant is
\be
ds^2_3 \sim \sin\theta (d\rho^2 + \frac{9 \rho^2}{6} d\Omega^2_{\rho} ) .
\end{equation}
This is in agreement with what we found before in
(\ref{eq:limbig}). It is interesting that the monopole and the string
have the same asympotic behaviour in the ALC case. This is possible
because the sphere becomes increasing cylindrical asymptotically, as was
commented in the previous section.
The asymptotic behaviour for the AC case may be found similarly.

\subsection{$\widetilde{\mathbb{C}}_7$ and $\mathbb{D}_7$}

The metrics for the $\widetilde{\mathbb{C}}_7$ and $\mathbb{D}_7$ families with a $U(1)$ symmetry both have the form
\be
ds_7^2 = dr^2 + a(r)^2 \left[ (\tilde{\sigma_1} + g(r) \sigma_1)^2 + (\tilde{\sigma_2} + g(r) \sigma_2)^2 \right]
+ b(r)^2 (\sigma_1^2 + \sigma_2^2)
+ c(r)^2 (\tilde{\sigma_3} + g_3(r) \sigma_3)^2
+ f(r)^2 \sigma_3^2 .
\end{equation}
The Taylor expansions about the origin are, for $\mathbb{D}_7$ \cite{cglp5, ab}
\bea
a(r) \sim \frac{r}{2} , & & b(r) \sim 1 - \frac{(q^2-2)r^2}{16}, \nonumber \\
c(r) \sim -\frac{r}{2} , & & f(r) \sim q + \frac{q^3r^2}{16}, \nonumber \\
g(r) \sim -\frac{a(r) f(r)}{2 b(r) c(r)} , & & g_3(r) \sim -1 + 2 g(r)^2  ,
\eea
where a trivial parameter has been fixed to one and $q$ is a free parameter. The AC solution
is recovered when $q=1$. There is an $S^3$ collapsing at the origin.

For $\widetilde{\mathbb{C}}_7$, the Taylor expansions about the origin are \cite{cglp6}
\bea
a(r) \sim 1+\frac{(4-c_0^2) r^2}{16} , & & b(r) \sim b_0 + \frac{(4 - 3b_0^2 c_0^2)r^2}{16 b_0}, \nonumber \\
c(r) \sim c_0 + \frac{(2+c_0^4) r^2}{4} , & & f(r) \sim (1+b_0^2)r, \nonumber \\
g(r) \sim -\frac{b_0 c_0 r}{2} , & & g_3(r) \sim b_0^2 - \frac{(1+b_0^2) r^2}{c_0^2} ,
\eea
setting a trivial parameter to one and with $b_0$ and $c_0$ as free parameters. There is an $S^1$ collapsing at the
origin.

Now compute the metrics near the origin as previously. There is a
subtlety which is that these families of metrics do not have a
$\mathbb{Z}_2$ symmetry interchanging $\sigma_i \leftrightarrow
\tilde{\sigma}_i$. Therefore we will get different metrics depending
on which of the $SU(2)$s we reduce on. To get a sensible metric in
four dimensions we should reduce on the copy that does not (partially) collapse at
the origin, which corresponds to the $\sigma_i$.
The result for $\mathbb{D}_7$ up to a constant, after putting $\rho = 2 r^{3/2}/3$,
\be
ds^2_3 \sim d\rho^2 + \frac{9 \rho^2}{16} (d\theta^2+\sin^2\theta d \psi^2) .
\end{equation}
This is exactly as for $\mathbb{B}_7$ in (\ref{eq:b7}).
Thus this family also give global monopoles. The result for $\widetilde{\mathbb{C}}_7$
up to a constant is
\be
ds^2_3 \sim \sin\theta \left[ dr^2 + d\theta^2 +
\left(\frac{1}{b_0^2} + 1\right)^2 r^2 d\psi^2 \right] .
\end{equation}
This metric is very similar to that for $\mathbb{C}_7$ in
(\ref{eq:c7}). It is cylindrical with a surfeit plane angle
that depends on $b_0$. There is therefore a natural interpretation as
a cosmic string. Also as for $\mathbb{C}_7$, spherical symmetry cannot
be restored for any value of $(b_0,c_0)$ and in particular it will
not be restored in the AC limit.

As in the previous subsection, we can also calculate the asymptotic
behaviour of the metric. We will do this for $\mathbb{D}_7$ in the ALC
case. From the first order equations for the radial functions \cite{cglp5} one
has that as $r\to\infty$
\bea
a(r)^2 \sim \frac{\sqrt{3}-1}{2}r^2 , & & b(r)^2 \sim \frac{\sqrt{3}-1}{2}r^2, \nonumber \\
c(r)^2 \sim (1-\sqrt{3})^2r^2 , & & f(r)^2 \sim k^2, \nonumber \\
g(r) \sim -\frac{a(r) f(r)}{2 b(r) c(r)} , & & g_3(r) \sim -1 + 2 g(r)^2  ,
\eea
where $k$ is a constant. Doing the dimensional reduction one obtains the asymptotic
metric up to a constant with $\rho=2 r^{7/2}/7$
\be
ds^2_3 \sim \sin\theta (d\rho^2 + \alpha \rho^2 d\widetilde{\Omega}^2_{\rho} ) ,
\end{equation}
with $\alpha = 4.48 \cdots$ and $d\widetilde{\Omega}^2_{\rho}$ is a metric on the
$S^2$ similar to (\ref{eq:sph}), increasingly cylindrical at infinity.

\section{Conclusion}

We have shown how one can obtain BPS monopoles in four dimensions
by dimensional reduction of special holonomy metrics of higher dimension. 
In particular,
we have considered seven dimensional Riemannian metrics of $G_2$ holonomy
extended trivially to eight dimensions by adding a time direction.

The $G_2$ metrics we considered all had $SU(2)\times SU(2)\times U(1)$
isometry group in the generic case. Four families of such metrics are known, denoted
$\mathbb{B}_7, \mathbb{C}_7, \widetilde{\mathbb{C}}_7, \mathbb{D}_7$. We
concentrated on the $\mathbb{B}_7$ case for which a couple of closed form
solutions are known. We reduced on $SU(2)\times U(1) \subset SU(2)\times SU(2)$
to get static four dimensional metrics with an $S^2$ factor and a $U(1)$ isometry group corresponding
to axial symmetry.

The $G_2$ metrics considered are generically asymptotically locally conical (ALC), with the $U(1)$ isometry
corresponding to a stabilised $S^1$. There is a special limiting case in each family of metrics in
which the metric becomes asymptotically conical (AC) with the $S^1$ blowing up at infinity.
For $\mathbb{B}_7$ and $\mathbb{D}_7$
the AC case has an enhancement of isometry to $SU(2)^3$. This is seen in the four dimensional
solution as an enhancement from axial to spherical symmetry. It was suggested that the preference
for ALC $G_2$ metrics could be understood from instabilities of certain spherically symmetric four dimensional
solutions with charged massive gauge fields under axially symmetric perturbations.

The families $\mathbb{B}_7$ and $\mathbb{D}_7$ have an $S^3$ that collapses at the origin and 
the four dimensional solution is spherically symmetric at the origin, with a conical singularity
at the origin hidden as a point of infinite redshift. These
are monopoles. The metric becomes increasingly cylindrical asymptotically.
For $\mathbb{C}_7$ and $\widetilde{\mathbb{C}}_7$, only an $S^1$ collapses at the origin and
this results in the four dimensional metric having axial symmetry at
the origin. These are cosmic strings. 

The $SU(2)\times U(1)$ gauge and scalar fields were also examined. It was shown that in the generic case
the $SU(2)$ gauge symmetry is completely broken except at infinity where it is broken to $U(1)$. This allows
a topological charge to be associated with the solution by constructing a map $S^2_{\infty} \to S^2$. This is not the
usual map because the Higgs fields are not in the usual
adjoint representation. There is a second topological charge from the $U(1)$ gauge symmetry.

\vspace{1cm}

Many thanks to Gary Gibbons for help and suggestions. The author is funded by the Sims scholarship.

\end{document}